\documentstyle[12pt]{article}
\begin{document}
\newcommand{\ba}{\begin{eqnarray}
\addtolength{\abovedisplayskip}{\extraspaces}
\addtolength{\belowdisplayskip}{\extraspaces}
\addtolength{\abovedisplayshortskip}{\extraspace}
\addtolength{\belowdisplayshortskip}{\extraspace}}
\newcommand{\one}{{\bf 1}}
\newcommand{\zbar}{\overline{z}}
\newcommand{\aalpha}{{\mbox{$\small \alpha$}}}
\newcommand{\ssigma}{{\mbox{$\small \sigma$}}}
\newcommand{\ea}{\end{eqnarray}}
\newcommand{\is}{& \!\! = \!\! &}
\addtolength{\oddsidemargin}{-1.2cm}
\addtolength{\topmargin}{-1cm}
\vspace{.5cm}
\addtolength{\baselineskip}{.5mm}
\input epsf
\newcommand{\newsubsection}[1]{
\vspace{1cm}
\pagebreak[3]
\addtocounter{subsection}{1}
\addcontentsline{toc}{subsection}{\protect
\numberline{\arabic{section}.\arabic{subsection}}{#1}}
\noindent{\large \bf \thesection.\thesubsection. #1}
\nopagebreak
\vspace{2mm}
\nopagebreak}
\newcommand{\ttau}{r}
\newcommand{\UU}{{\cal W}}
\newcommand{\vev}[1]{\langle #1 \rangle}
\def\mapright#1{\!\!\!\smash{
\mathop{\longrightarrow}\limits^{#1}}\!\!\!}
\newcommand{\bigoint}{\displaystyle \oint}
\newlength{\extraspace}
\setlength{\extraspace}{2mm}
\newlength{\extraspaces}
\setlength{\extraspaces}{2.5mm}
\newcounter{dummy}
\newcommand{\be}{\begin{equation}
\addtolength{\abovedisplayskip}{\extraspaces}
\addtolength{\belowdisplayskip}{\extraspaces}
\addtolength{\abovedisplayshortskip}{\extraspace}
\addtolength{\belowdisplayshortskip}{\extraspace}}
\newcommand{\ee}{\end{equation}}
%
\newcommand{\twomatrix}[4]{{\left(\begin{array}{cc}#1 & #2\\
#3 & #4 \end{array}\right)}}
\newcommand{\twomatrixd}[4]{{\left(\begin{array}{cc}
\displaystyle #1 & \displaystyle #2\\[2mm]
\displaystyle  #3  & \displaystyle #4 \end{array}\right)}}
\newcommand{\ie}{{\it i.e.\ }}
\newcommand{\half}{{\textstyle{1\over 2}}}
\newcommand{\brst}{{\sc BRST}\ }
\newcommand{\pl}{{\strut +}}
\newcommand{\mi}{{\strut -}}
\newcommand{\Q}{{\bf\sc Q\, }}
\newcommand{\V}{V}
\newcommand{\I}{{\cal I}}
\newcommand{\D}{{\cal D}}
\newcommand{\T}{{\cal T}}
\newcommand{\1}{{\it 1}}
\newcommand{\figuurb}[3]{
\hspace{-3mm}{\it }\ 
\begin{figure}[ht]\begin{center}
\leavevmode\hbox{\epsfxsize=#2 \epsffile{#1.eps}}\\[3mm]
\parbox{11cm}{
\it #3}
\end{center} \end{figure}\hspace{-5.5mm}}
\newcommand{\fig}{{\it}}
\newcommand{\baa}{\begin{eqnarray}
\addtolength{\abovedisplayskip}{\extraspaces}
\addtolength{\belowdisplayskip}{\extraspaces}
\addtolength{\abovedisplayshortskip}{\extraspace}
\addtolength{\belowdisplayshortskip}{\extraspace}}
\newcommand{\eaa}{\end{eqnarray}}
\newcommand{\Z}{{\bf Z}}           
\newcommand{\hf}{{\textstyle{1\over 2}}}
\newcommand{\SF}{{\bf S}^5}
\newcommand{\RF}{{\bf R}^4}
\newcommand{\ra}{\rightarrow}
\newcommand{\la}{\leftarrow}
\newcommand{\delbar}{\overline{\partial}}
\newcommand{\R}{{\bf R}}
\renewcommand{\Im}{{\rm Im\,}}
\def\a{\alpha} 
\def\b{\beta} 
\newcommand{\ppar}{{{}_{\!\!{}/\!/}}}
\newcommand{\del}{\partial}
\def\e{\epsilon}
\newcommand{\sbset}{\subseteq }
\addtolength{\baselineskip}{.5mm}
\renewcommand{\thesubsection}{\arabic{subsection}}
\newcommand{\figuur}[3]{
\begin{figure}[t]\begin{center}
\leavevmode\hbox{\epsfxsize=#2 \epsffile{#1.eps}}\\[3mm]
\bigskip
\parbox{15.5cm}{\small \ 
\it #3}
\end{center} \end{figure}\hspace{-1.5mm}}

\begin{titlepage}
\begin{center}

{\hbox to\hsize{
\hfill PUPT-1923}}

\bigskip
\bigskip

\vspace{6\baselineskip}

{\large \bf A Note on Warped String Compactification}

\bigskip


\bigskip
\bigskip
{  Chang S. Chan${}^1$, Percy L. Paul${}^2$ and Herman
Verlinde${}^1$}\\[1cm]

${}^1$ { \it Joseph Henry Laboratories,
Princeton University, Princeton NJ 08544}\\[.5cm]

${}^2$ {\it National Research Council,
100 Sussex Drive, Ottawa Ontario K1A OR6}

\vspace*{1.8cm}

{\bf Abstract}\\

\end{center}
\noindent
We give a short review of a large class of warped string geometries,
obtained via F-theory compactified on Calabi-Yau fourfolds, that upon
reduction to 5 dimensions give consistent supersymmetric
realizations of the RS compactification scenario.

\end{titlepage}

\subsection{Introduction}

\smallskip

This note is intended to clarify the realization and
interpretation of the Randall-Sundrum compactification scenario within
string theory. In the model
of \cite{rs}, our 4-d world is extended with an extra direction $r$ 
to a 5-d space-time with the warped metric 
\be
\label{een}
ds^2 \, = \,  e^{2\ssigma(r)}\,  \eta_{\mu\nu} \, dx^\mu dx^\nu + dr^2
\ee
with $\sigma(r) = -k|r|$. Even while the range of $r$ is infinite, 
the warped form of the metric ensures that the effective
volume of the extra direction is finite. As a result, matter
particles sufficiently close to the domain wall region near $r= 0$ 
will experience ordinary 4-d gravity at long distances \cite{rs}.

\smallskip

At first sight, this proposal seems like a rather drastic departure
from the more conventional Kaluza-Klein framework. Indeed, in most
works on string compactifications thus far, the four uncompactified
directions and the compact manifold are assumed to form a simple
direct product. Although it was realized for a long time that this
basic KK set-up can be generalized to include the possibility of
warped products, the physics of these more general scenarios is still
largely unexplored.

\smallskip

A second important ingredient of the RS-scenario 
is that part or
all of the observable matter may be thought of as confined to a 4-d
sub-manifold of the higher dimensional space-time. 
A concrete theoretical realization of such
world-branes are the D3-branes of IIB string theory, which confine
open strings to their world-volume. D3-branes, however, do not bind 
4-d gravity. Possible supersymmetric realizations of the
Planck-brane, located around $r=0$ in (\ref{een}), are therefore
rather expected to be found in the form of domain wall type configurations,
or possible stringy generalizations thereof. Various attempts have
been made to find smooth domain wall solutions of this type within
5-d gauged supergravity, but thus far without real success 
\cite{kallosh} \cite{cvetic}.

\smallskip

There are several reasons for why this is indeed a hard problem.  
Even for a given compactification from 10 dimensions, it is an elaborate
task to derive the dimensionally reduced theory. Thus far this has been 
done only for reductions over rather special symmetric 5-manifolds 
$K_5$ such as $S^5$ or $S^5/Z_2$, etc, and/or for special theories with 
extended supersymmetry. However, while it seems feasible to classify 
the possible types of supersymmetric solutions for each of these special 
dimensional reductions, there is no guarantee that they provide a 
general enough framework.

\smallskip

Instead of following the above procedure of $(1)$ performing some
special dimensional reduction to 5 dimensions and $(2)$ looking for RS
domain wall type solutions, it seems more practical to reverse the two
steps. Since the scalar fields $\phi^a$ arise as moduli of some
internal 5-d compact space $K_5$, {\it any} domain wall solution in
5-d gauged supergravity describes (upon lifting it back up to
10-dimensions) some specific warped compactification of the 10-d
theory. It will therefore be much more general -- and also easier --
to {\it first} (a) identify a general class of warped
compactifications of the 10-dimensional theory, and {\it then} (b)
perform the same type of dimensional reduction from 10 to 5
dimensions.  In the end, one can then hope to identify a class of 10-d 
geometries for which the resulting dimensionally reduced
solution has all the required properties.

\figuur{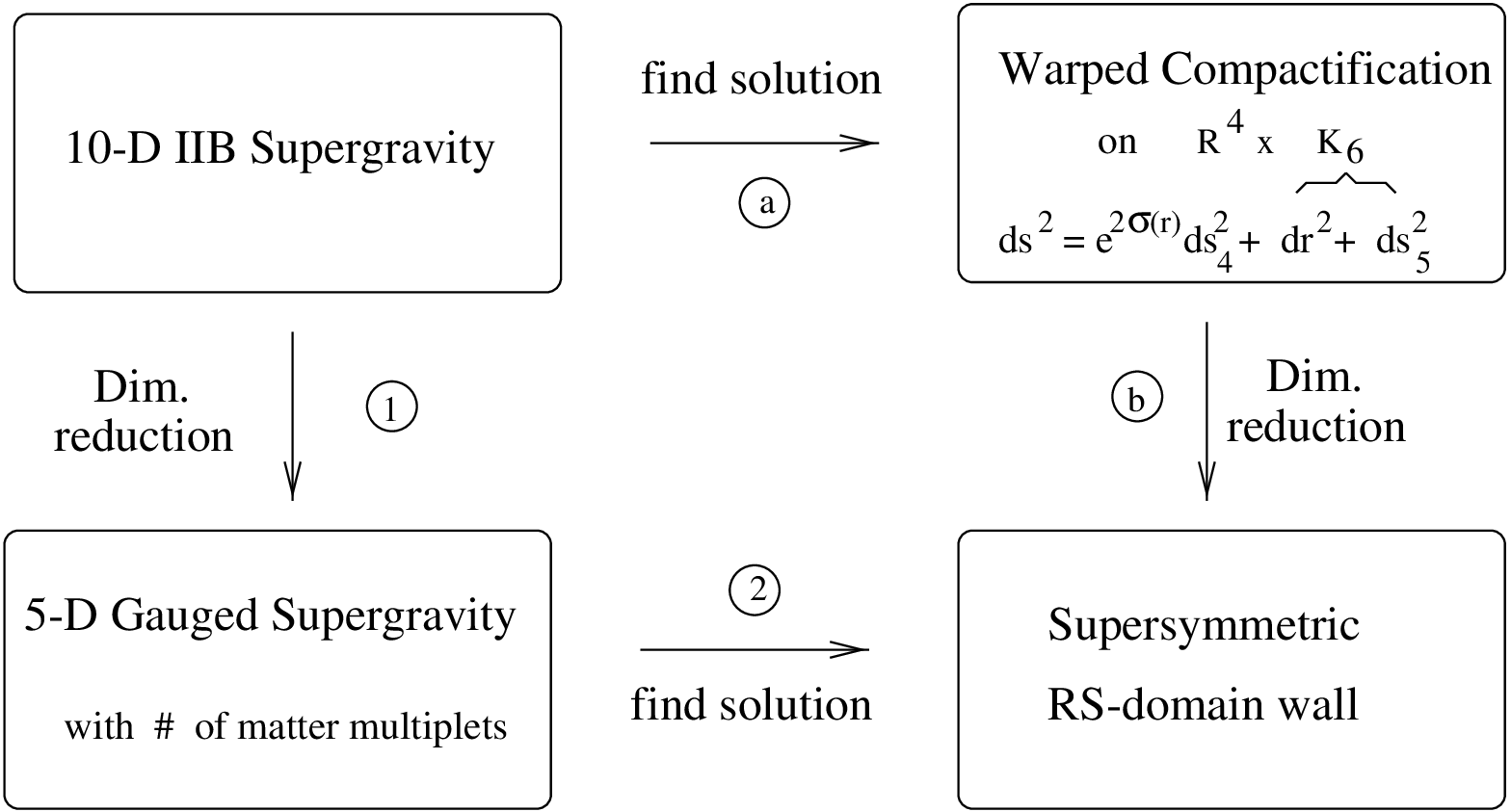}{10cm}{{\bf Fig 1.} \ 
To identify supersymmetric 
RS-type geometries, we will follow the route {\rm IIB
--$>$ (a) --$>$ (b) --$>$ RS.} It is still an open problem to
find a direct construction of these geometries via the other route.\\[-9mm]}

\vspace{-6mm}

\smallskip

As will be described below, such a class of warped IIB geometries
indeed exists in the form of quite generic F-theory compactifications 
on Calabi-Yau four-folds.\footnote{Another special realisation 
of an RS geometry in terms of a toroidal type IIB orientifold 
compactification has been described in \cite{hv}.}
These have been studied in some detail in
the recent literature -- a list of references include
\cite{becker} \cite{DRS}\cite{wittenf}\cite{svw} and \cite{gvw} -- and
indeed none of our equations will be new. Given the current interest
in the subject, however, it seems useful to collect some of the known
facts about these compactifications, since it has not been generally
appreciated that supersymmetric RS-geometries indeed exist in
string theory, and furthermore that they are in fact quite generic.

\smallskip

Since all derivations are contained in existing papers, we will here
only present the general form of the compactification geometry without
any proof that it is really a supersymmetric solution to the 10-d
equations of motion. This proof can however be quite directly
extracted from the literature, in particular from the very clear 
discussion by Becker and Becker \cite{becker}. Their analysis was
done in the context of M-theory compactifications on C-Y four-folds. 
It can however be straightforwardly translated to the F-theory
context by performing the T-duality transformation outlined in
\cite{wittenf}. An explicit example of this T-duality transformation 
is discussed in \cite{DRS}.

\smallskip
 
Although the 10-d perspective will allow us
to identify a large class of RS-type compactification geometries,
their geometrical structure is rather involved. It is therefore not
easy to {\it explicitly} perform the dimensional reduction of these
solutions to 5-dimensions. We will nonetheless attempt to make this
5-d perspective as transparent as possible. In particular we
will show that they indeed give rise to a 5-d metric of the
generic form (\ref{een}).


\subsection{Warped Compactification in F-theory}

\newcommand{\MM}{{\raisebox{-.4ex}{\mbox{${\rm {}^{{}_M}}$}}}}
\newcommand{\NN}{{\raisebox{-.4ex}{\mbox{${\rm {}^{{}_N}}$}}}}

\noindent
F-theory is a geometric language for describing compactifications of
type IIB string theory, in which the expectation values of the dilaton
and axion fields are allowed to vary non-trivially along the
compactification manifold \cite{vafa}.  Compactifications of F-theory
down to four-dimensions are specified by means of a Calabi-Yau
four-fold that admit an elliptic fibration with a section. In other
words, these are 8 dimensional compact manifolds $K_8$ that {\it
locally} look like a product of a complex three-fold $K_6$ times a
two-torus $T^2$. The two-torus will be taken to shrink to zero
size. It can however be taken to change its shape when moving along
the base $K_6$. In particular, it can have non-trivial monodromies
around singular co-dimension 2 loci inside the $K_6$, where the 
elliptic fiber degenerates.

\smallskip

The four-fold $K_8$ is not the actual compactification geometry;
rather it gives an economical way to characterize the compactification
geometry as well as the expectation values of other fields.
Moreover, due to the special geometric properties of the
Calabi-Yau four-fold $K_8$ -- vanishing first Chern class and $SU(4)$
holonomy -- the associated IIB background by construction will
preserve 4-d supersymmetry, at least at the classical and perturbative
level.

\smallskip

The warped geometry of this type of F-theory compactifications has
been derived in \cite{DRS}, by direct translation of the M-theory
analysis of \cite{becker}. The full solution for the
10-dimensional IIB string metric, in the string frame, takes the form
\be
\label{twarp}
ds_{{}_{IIB}}^2 
= e^{2\aalpha(y)} \eta_{\mu\nu} dx^{\mu} dx^\nu\, +\, 
e^{-2\aalpha(y)} \, g_{\MM\NN}(y) dy^\MM dy^\NN 
\ee
where $g_{\MM\NN}$ denotes the metric on the 6-dimensional base-manifold
$K_6$. The shape of the warp-factor $e^{2\alpha}$ will depend on the
detailed geometry of the CY four-fold $K_8$, as well as on other data
such as the possible non-zero expectation values of other fields
and the locations of the possible D-branes. 

\smallskip

Besides the ten-dimensional space-time metric, the fields that can take 
non-trivial expectation values are the following:
\begin{center}
\parbox{15cm}{
(i)\hspace{4.4mm} the dilaton field \hspace{5.5cm} $\phi$\\[1mm] 
(ii)\hspace{3.3mm} the RR-scalar or axion field \hspace{3.4cm} 
$\widetilde{\phi}$\\[1mm]
(iii) \hspace{1.1mm} the NSNS 3-form field strength \hspace{2.7cm}
$H^{NS}$\\[1mm]
(iv) \hspace{1.4mm} the RR 3-form field strength\hspace{3.3cm} $H^R$\\[1mm]
(v)\hspace{3.8mm} the RR 5-form field strength \hspace{3.2cm} 
$F^{(5)}$}
\end{center}

\noindent
The expectation values of all these fields can all be 
conveniently characterized in terms of the geometry of $K_8$.

\smallskip

In the type IIB theory, the modulus $\tau$ of the elliptic fibration,
the shape of the two-torus inside the $K_8$, describes the variation
along the 6-d base manifold $K_6$ of the dilaton and axion fields,
$\phi$ and $\widetilde{\phi}$, via the identification
\be
\label{tauF}
\tau = \widetilde{\phi} + {i e^{-\phi}}.
\ee
As mentioned above, a key feature of F-theory is that this modulus
in general has non-trivial monodromies around 4-d
submanifolds inside $K_6$. These 4-d sub-manifolds are associated with
the locations of D7-branes,
of which the remaining 3+1-dimensions span the uncompactified
space-time directions. In going around one of the D7-branes,
the modulus field $\tau$ can pick up an $SL(2,\Z)$ monodromy 
\be
\label{slto}
\tau \rightarrow \frac{a\tau +b}{c\tau +d},
\ee
which leaves the geometric shape of the two-torus fibre inside $K_8$
invariant, but nonetheless via (\ref{tauF})
amounts to a non-trivial duality transformation of the IIB string
theory. We thus notice that the dilaton and axion field are not smooth
single-valued functions, but instead are multi-valued with branch cut
singularities at the locations of the D7-branes. The full 
non-perturbative string theory, however, is expected to be well-behaved
everywhere.

\smallskip

For the following, it 
will be convenient to combine the NSNS and RR three-from field strengths,
$H^{NS}$ and $H^R$, of the IIB supergravity into a single four-form 
field-strength $G$ on $K_8$ as
follows \cite{gvw}. Let $z$ and $\zbar$ denote the coordinates along 
the $T^2$ fiber. Then we can write
\be
G \, =\, {\pi \over i \tau_2  }( \, H \wedge d\zbar \, - \, 
\overline{H} \wedge dz\, )
\ee
\be
H = H^{R} \! - \tau H^{NS} \; ; \qquad 
\overline{H} = H^{R} \! - \overline{\tau} H^{NS}
\ee
For supersymmetric configurations, $H$ defines an integral harmonic
(1,2)-form on $K_6$ satisfying $H\wedge k = 0$ with $k$ the K\"ahler
class of $K_6$ \cite{gvw}. It transforms under the $SL(2,Z)$ monodromy
transformations (\ref{slto}) around the seven-branes as $H\rightarrow
H/(c\tau +d)$.  The field-strength $G$ is invariant under these
transformations.

\smallskip

An important aspect of F-theory compactifications is that they
typically carry, via their non-trivial topology, an effective total
D3-brane charge. The value of this charge is proportional to the
Euler characteristic $\chi(K_8)$ of the original Calabi-Yau four-fold
$K_8$. Here $\chi(K_8)$ is defined via
\be
\label{chi}
{1\over 24} \chi(K_8) = \int_{K_8} I_8(R)
\ee 
where
\be
I_8(R) = {1\over 192} \Bigl({\rm tr}{R}^4 - {1\over 4} (
{\rm tr}{R^2})^2\Bigr)
\ee
with $R$ the curvature two-form on $K_8$. Global tadpole cancellation,
or conservation of the RR 5-form flux, requires that this charge must
be canceled by other sources.  These other sources come from
possible non-zero fluxes of the NSNS or RR two-form fields, or 
from the explicit insertion of $N$ D3-brane
world branes, that is, D3-branes that span the 3+1-d uncompactified
world but are localized as point-like objects inside the $K_6$. 
The number of such D3-branes is therefore not free, but completely
determined via charge conservation. This 
global tadpole cancellation relation reads
\be
\label{ncount}
N = {1\over 24} \chi(K_8)  - {1\over 8\pi^2} \int_{K_8} G \wedge G
\ee
Depending on the topology of $K_8$,  
$N$ can reach values of up to $10^3$ or larger. An example with $N=972$, 
mentioned in \cite{svw}, is provided by an elliptically fibered CY 
four-fold over ${\bf P}^3$. The Euler number $\chi(K_8)$ can be 
non-zero only if $K_6$ has a non-vanishing first Chern class, that is,
provided the F-theory compactification makes use of a non-zero
number of D7-branes.

\smallskip

The equation of motion for the warp factor $e^{2\aalpha}$ obtained in
\cite{becker} and \cite{DRS} reads as follows
\be
\label{delta}
\Delta^{(8)}\, e^{-4\aalpha} \, = \,  * \; 4\pi^2 \left\{  \, I_8(R) \, 
- {1\over 8\pi^2} G \wedge G - \sum_{i=1}^N \delta^{(8)}(y\! -\! y_i) \,
\right\}
\ee
where $\Delta^{(8)}$ denotes the Laplacian and $*$ the Hodge star 
on $K_8$. The points $y=y_i$ correspond to the location of the $N$
D3-branes. Here, following
\cite{DRS}, we have written the equation on the full 8-d
manifold $K_8$, even though in F-theory the elliptic fiber $T^2$ inside 
$K_8$ has been shrunk to zero size. In this limit, the solution for 
$\alpha$ obtained via (\ref{delta}) only depends on the 6 
coordinates $y^\MM$ on $K_6$. Alternatively, using the analysis of 
\cite{svw}, one may also first reduce the right-hand side to $K_6$, 
via integration over the $T^2$ fiber, and then solve the reduced 
equation to obtain the function $e^{-4\aalpha}$ directly on $K_6$.

\smallskip

Finally, there is also an non-trivial
expectation value for the self-dual RR five-form field strength,
equal to \cite{becker} \cite{DRS}
\be
F_{\mu\nu\lambda\sigma\MM} = \epsilon_{\mu\nu\lambda\sigma}
\partial_{\MM} e^{-4\alpha} .
\ee
We note that via (\ref{delta}) the D3-branes
indeed form a source for this field strength, but that via (\ref{ncount}) 
the total charge adds up to zero.

\figuur{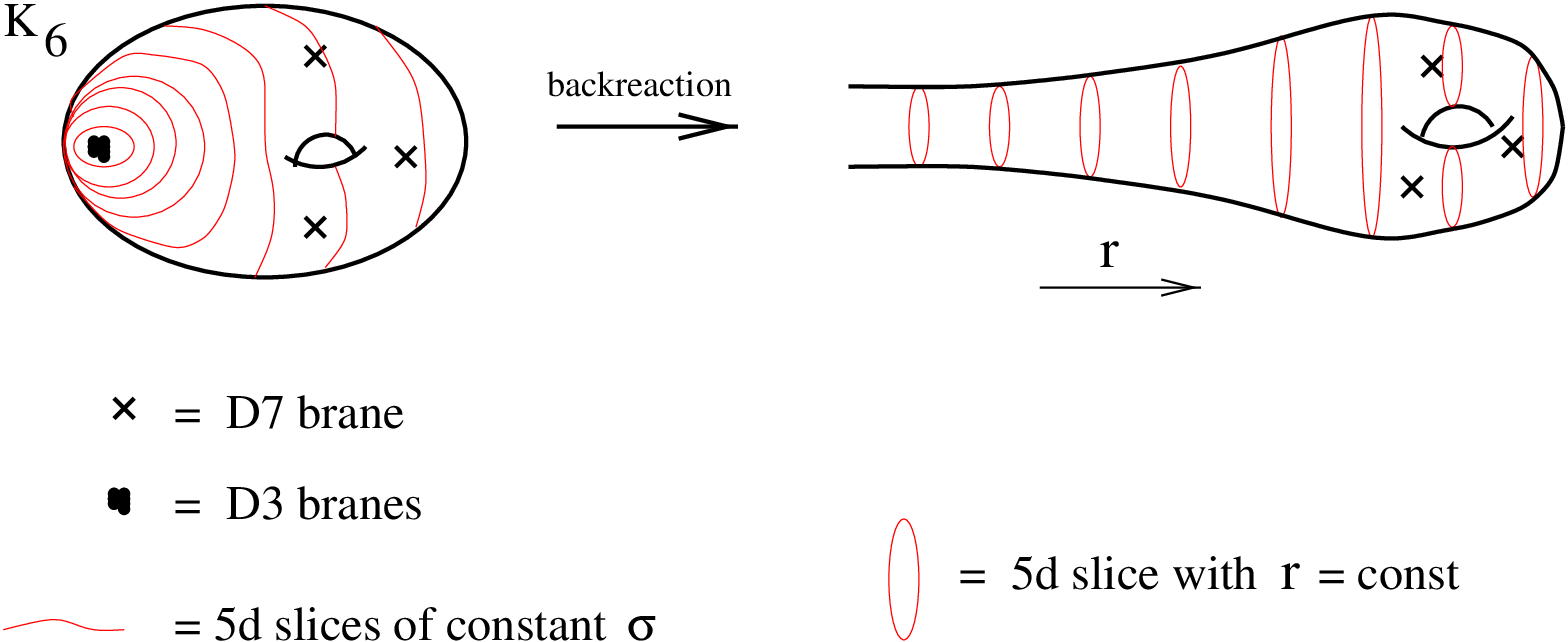}{11cm}{{\bf Fig 2.}\ The contours with constant warp
factor $e^{2\sigma}$ define a particular slicing of the 6-dimensional
compactification manifold $K_6$, which can be used to
represent $K_6$ as a one parameter flow along $r$ of five-manifolds
$K_5$. Upon dimensional reduction to 5 dimension, 
this geometry describes a one-sided RS-domain wall solution.  
\\[-7mm]}

\vspace{-5mm}

\subsection{Shape of the warp factor}

Let us summarize. Starting from the elliptically 
fibered  CY four-fold $K_8$ we can extract a complete 
characterization of the warped compactification. First, 
since $K_8 \simeq K_6 \times T^2$, we obtain the metric 
$g_{\MM\NN}$ on the base $K_6$, as well as the dilaton and
axion via (\ref{tauF}). We then deduce the form of the warp 
factor $e^{2\aalpha}$ from (\ref{delta}), which incorporates
the complete backreaction due to the $G$-flux and D3-branes.
Finally from (\ref{twarp}), we obtain the actual compactification 
geometry. Note that, as indicated in fig 2, the rescaling by 
$e^{-2\aalpha}$ of $g_{\MM\NN}$ in (\ref{twarp}) may have 
a drastic effect on the shape of the compactification manifold, which 
indeed may look quite different from that of the original $K_6$. 
In particular, it is possible that near the 
locations of the D3-branes one of the internal directions may 
become non-compact.

\smallskip

We may formally solve the equation (\ref{delta}) via
\be
\label{sol}
e^{-4\aalpha(y)} = e^{-4\aalpha_0} + 
4\pi^2 \int \! d^8y' \sqrt{\, g\, } \, {\cal G}(y,y') 
\Bigl[\, I_8(R(y'))  - {1\over 8\pi^2} G \wedge G
- \sum_{i=1}^N \delta^{(8)}(y'-y_i) \, \Bigr]
\ee
where ${\cal G}(y,y')$ denotes the Green function for $\Delta^{(8)}$. 
The term $e^{-4\aalpha_0}$ parametrizes
the constant zero mode of $e^{-4\aalpha}$, which is not
fixed by eqn (\ref{delta}). Note that for
$e^{-4\aalpha}$ to be everywhere positive, this 
constant $e^{-4\aalpha_0}$ can not be arbitrarily small, 
since the second term on the r.h.s. of (\ref{sol})
can become negative. This implies that
the warped 6-geometry automatically has a {\it minimal} volume\footnote{We 
thank Sav Sethi for bringing this feature to our attention.}. 
\smallskip

An interesting limiting case is when
all $D3$ branes are concentrated in one point, say $y=y_0$.
Close to this point, the warp function $\alpha(y)$ reduces to 
\be
\aalpha(y) \simeq \log |y-y_0| \, + \, {\rm const.}
\ee 
Via (\ref{twarp}) this describes the familiar semi-infinite 
near-horizon geometry of $N$ D3-branes: 
$AdS_5 \times S^5$  with radius $R= {}^4\!\! \sqrt{4\pi N g_s}$.
(See fig 2.) Although the radial
$AdS_5$ coordinate $r \simeq - R \log|y-y_0|$ runs over semi-infinite 
range, the compactification geometry (\ref{twarp})
still gives rise to a 4-d Einstein action with a
{\it finite} 4-d Newton constant $1/\ell_4^2$ equal to
\be 
\label{plank}
{1\over (\ell_{4})^2}\, = \,{ 1 \, 
\over (\ell_{{}_{\! 10}})^8 }
\int_{K_6} \!\! \sqrt{\, g\,} \,e^{2\phi-4\aalpha} 
\ee
with $\ell_{10}$ the 10-d Planck length.

\subsection{Reduction to 5 dimensions.}

\noindent
We would now like to show that these F-theory compactifications,
upon performing a suitable dimensional reduction to five dimensions,
reduce to supersymmetric RS domain wall solutions. 
To this end we will look for a specific coordinate system
\be
y^\MM = (y^m , r)
\ee
where $m$ now runs over 5 values, such that the 10-d metric
takes the following form
\be
\label{rsmet}
ds_{RS}^2 \; = \;
e^{2\sigma(r)}  \eta_{\mu\nu} \, dx^\mu dx^\nu + dr^2
+ h_{mn}(y,r) dy^m dy^n
\ee
where $ds_{RS}^2$ related to the original IIB string metric (\ref{twarp})
via the rescaling
\be
\label{rescale}
ds^2_{RS} 
=\, {e^{\phi/2} (V_5)^{1/4}}
\; ds^2_{{}_{IIB}}\ 
 \qquad \qquad 
V_5 = {1\over (\ell_{10})^5} \, \int_{K_5}\! \sqrt{\, h}
\ee
Here the prefactor in (\ref{rescale}) is chosen such that $ds^2_{RS}$
is the metric in the 5-d Einstein frame (where we have taken the 5-d
Planck length $\ell_5$ equal to the 10-d one). The 5-d slices of
constant $r$ define 5-d submanifolds $K_5$, on which $2\alpha(y) +
\phi(y)/2 =$ constant; this correspondence guarantees that the
warp-factor $e^{2\sigma}$ in $ds^2_{RS}$ just depends on $r$ and not
on the remaining $y^m$'s.

\smallskip

In this way we indeed obtain a solution that from the 5-d perspective
looks just like an RS-type warped geometry. For large negative $r$,
close to the D3-branes, the warp factor behaves like $e^{2\sigma}
\simeq e^{-2 |r|/R}$ with $R$ the AdS-radius of the $N$ D3-brane
solution. On the other end, somewhere outside the throat
region of the AdS-tube, near the `equator' of the $K_6$, the warp 
factor $e^{2\sigma(r)}$ reaches some maximal value.
Eventually, there is a boundary value for the coordinate $r$, which we
can take to be $r=0$, at which the transverse 5-manifold $K_5$ shrinks
to zero size.  However, as is clear from fig 2, this does not
correspond to any singularity, but just to a smooth cap closing off
the 6-d manifold $K_6$. This fact that $r$ takes a maximal value $r=0$
implies that the AdS-space is indeed compactified, in the sense that,
relative to the 4-d space-time, it has a finite volume. It therefore
produces to a finite 4-d Newton constant equal to (cf \cite{rs})
\be
{1\over (\ell_4)^2} = {1\over (\ell_5)^3} \;
\int\limits_{-\infty}^0 \! dr \, e^{2\sigma(r)}.
\ee
It is easily verified that via (\ref{rescale}) and (\ref{rsmet}),
this result coincides with the value (\ref{plank}) found via
direct reduction from 10 dimensions.

\smallskip

Upon dimensional reduction, the metric $h_{mn}$ of the internal $K_5$,
as well as other fields such as the dilaton, all reduce to scalar
fields that provide the matter multiplets of the 5-d supergravity. All
these fields vary with the radial coordinate $r$, and since this
radial flow is supersymmetric, it should in principle be described as
some gradient flow driven by some appropriate superpotential. 
However, due to the rather complex geometrical structure of the
typical F-theory compactification, it unfortunately seems impossibly
hard to find the explicit form of this potential.

\figuur{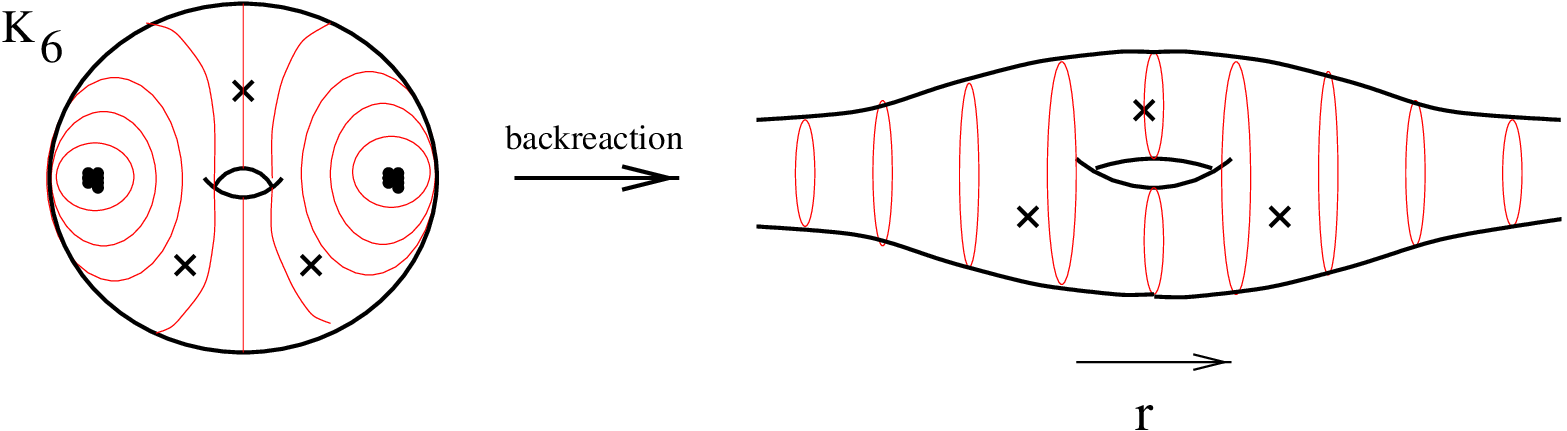}{10cm}{{\bf Fig 3.} \ An RS domain wall in between 
two AdS-type regions can be obtained by starting with a ${\bf Z}_2$ 
symmetric 6-manifold $K_6$, in which the D3-branes are located at 
opposite image points under the ${\bf Z}_2$.\\[-7mm]}

\subsection{Discussion}

In this note we have summarized the geometrical description of warped
F-theory compactifications, and shown they can be used to obtain
geometries very analogous to the RS-scenario.  Due to the presence of
the D7-branes in the F-theory geometry, the solutions are not
completely smooth: the dilaton and axion field have isolated
branch cut singularities at the D7-brane loci around which $\tau$
is multi-valued. The string theory, however, is well-defined in this
background.

\smallskip

Our description of the solutions makes clear that the internal
structure of the RS Planck brane is not that of an actual brane, but
rather that of a compactification geometry (which however also contains
the D7-branes).  Consequently, the localized graviton zero mode 
of \cite{rs} is just
the standard KK zero mode of the 10-d metric; because its
wave-function is sharply peaked near the wall region, where the warp
factor $e^{2\ssigma}$ is maximal, the 4-d graviton indeed {\it looks
like} some bound state. From the higher dimensional viewpoint, however,
this RS-localization of gravity is not a new phenomenon.

\smallskip

The most interesting aspect of these warped string compactifications,
is that there is no clear distinction between the low energy and extra
dimensional physics.  Kaluza-Klein excitations, when localized far
inside the AdS region can describe particles with masses much smaller
than the inverse size of the original $K_6$.  While these particles
naively look like new degrees of freedom arising from the presence of
extra dimensions, the holographic AdS/CFT correspondence \cite{adscft}
tells us that they
are in fact localized excitations of the low energy gauge theory.  
The same holds for string excited modes in this region. 
Hence via the holographic identification of the RG
scale with a {\it real} extra dimension, the two usually separate
stages of dimensional and low energy reduction should now be combined
into one single procedure.

\smallskip

Finally, as a closely related point, we need to emphasize that the
solutions as given here are generically unstable against small
perturbations.  The best way to understand this instability is via the
RG language: in general there will exist relevant operators whose
couplings, once turned on in the UV region, will quickly grow and
typically produce some singularity that effectively closes off the
AdS-tube \cite{ps}. In our way of obtaining the solutions, we did not
immediately notice this instability, because we required that the
original $K_6$ geometry and all other fields, except for the
warp-factor and the 5-form flux, are smooth at the locations of the
D3-branes. This requirement is special, however, and we should allow
for deformations that may spoil this property.

\smallskip 
 
In order to ensure that the perturbed geometry contains a substantial
intermediate AdS-like region, we either need to fine-tune the UV
initial conditions or introduce a symmetry that eliminates these
unstable modes. In RG language, this means that the dual 4-d field
theory should be made approximately conformal invariant over a large
range of scales, separating the Planck scale from the scale set by the
non-trivial gauge dynamics. Via the AdS/CFT dictionary, the problem of
realizing an RS-geometry in string theory therefore is reduced back to
the original problem it was designed to solve, namely how to generate
a large gauge hierarchy. Or stated in more positive terms, in
searching for realistic string compactification scenarios, the
observed gauge hierarchy can be viewed as an indication that warped
geometries of this type deserve serious attention.

\bigskip

\noindent
{\sc Acknowledgements} \\ 
This work is supported by NSF-grant
98-02484.  C.S.C is supported in part by an NSF Graduate Research Fellowship. 
We would like to thank M. Berkooz, M. Cvetic, R. Dijkgraaf, M. Gremm,
R. Kallosh, L. Randall,  S. Sethi and E. Verlinde for helpful discussions.

\bigskip

\bigskip
\renewcommand{\Large}{\large}

\end{document}